\documentclass{PoS}
\newcommand{\bb}{\begin{equation}}
\newcommand{\ee}{\end{equation}}
\newcommand{\eqb}{\begin{eqnarray}}
\newcommand{\eqf}{\end{eqnarray}}
\newcommand{\hs}{/\kern-.52em h}

\newcommand{\id}
 {i\kern.06em\hbox{\raise.25ex\hbox{$/$}\kern-.60em$\partial$}}
\newcommand{\as}{/\kern-.68em A}
\newcommand{\Ds}{/\kern-.69em D}

\newcommand{\ks}{/\kern-.67em k}
\newcommand{\Ps}{/\kern-.65em p}

\newcommand{\bs}{/\kern-.52em b}
\newcommand{\qs}{/\kern-.52em s}

\def\p{{\bf p}}
\def\x{{\bf x}}
\def\bn{{\boldsymbol{ \nabla}}}
\def\bth{{\boldsymbol{ \theta}}}
\usepackage{amsmath}
\def\B{{\bf{ B}}}
\title{CPT/Lorentz Invariance Violation and Quantum Field Theory}

\ShortTitle{Noncommutative fields}

\author{\speaker{Jorge Gamboa}\\
Departamento de F\'{\i}sica, Universidad de Santiago de Chile\\
Casilla 307, Santiago 2, Chile\\
E-mail: \email{jgamboa@lauca.usach.cl}}

\author{Paola Arias\\
Departamento de F\'{\i}sica, Universidad de Santiago de Chile\\
Casilla 307, Santiago 2, Chile\\
E-mail: \email{paola.arias@gmail.com}}

\author{Ashok Das\\
Department of Physics and Astronomy, University of
  Rochester, Rochester, NY 14627-0171, USA \\
  and
  \\
 Saha Institute of Nuclear Physics, 1/AF Bidhannagar, Calcutta 700064, India.
  \\
E-mail: \email{das@pas.rochester.edu}}

\author{Justo Lopez-Sarrion\\
Department of Physics, City College of  CUNY\\
New York, NY 10031, USA\\
E-mail: \email{justinux75@gmail.com}}

\author{Fernando Mendez\\
Departamento de F\'{\i}sica, Universidad de Santiago de Chile\\
Casilla 307, Santiago 2, Chile\\
  E-mail: \email{ fmendez@lauca.usach.cl}}

\abstract{Analogies between the noncommutative harmonic oscillator and
noncommutative fields are analyzed. Following this analogy we
construct examples of quantum fields theories with explicit CPT and
Lorentz symmetry breaking. Some applications to baryogenesis and
neutrino oscillation are also discussed.
}

\FullConference{Fifth International Conference on Mathematical Methods in Physics --- IC2006\\
         April 24-28 2006\\
         Centro Brasilerio de Pesquisas Fisicas, Rio de Janeiro, Brazil}

\begin{document}

\section{Introduction}

The Lorentz and CPT symmetries are fundamental cornerstones in the
twentieth century for which there are not clear indications if
persists at very high energies. By other hand, from effective
theories point of view, it seems natural to think that the presently
quantum field theories (QFT) and the symmetry principles they are
based on, remain valid just for a given range of energies beyond
which, possibly new and unexpected phenomena could emerge. For
example, the relativistic invariance itself could be broken or
deformed \cite{effective}.

If QFT describes fundamental interactions for any energy range,
then it seems natural to think that any QFT --seen as effective
theory-- must incorporate two energy scales, namely, the infrared and
ultraviolet ones. Both scales might give rise to non conventional
implications.

 There are at least two examples of the previous idea, where the
 infrared scale can be very important. The first one, is the physics
 involved in the infrared sector of QED where still several technical
 aspects need to be understood as well as many conceptual
 problems still remain open \cite{yennie}.

The infrared sector of QED is the natural link between quantum field
theory and quantum mechanics and then we ask which are the problems and how
can we understand the physics in this interface?, what are the most
convenient approximation criteria?  In spite of many efforts
performed during the fifties and sixties this problem still have not
been clarified.

In the same context, another important example is QCD where  the physical picture
in the infrared limit is nontrivial because,  at low
energy, the theory is nonperturbative and phenomena such as
confinement or  hadronization should be solved using new methods beyond
perturbation theory.

In this paper we would like to report our previous   results on the idea
of how the Lorentz symmetry could be broken in a QFT and also to point
out new progress in the application of this to neutrino physics.

In order to expose our results, we will start in section II describing
noncommutative mechanics and the harmonic oscillator in order to introduce, in the
same section,  the notion  of  noncommutative scalar and gague fields. In section III we
we will report our recent progress in neutrino physics when  noncommutative
fermion fields are introduced. Section IV is devoted to resume the results of
previous ideas when they are applied to the study of early universe physics. In the last
section, conclusion and outlook is presented.

\section{Noncommutative quantum mechanics and noncommutative
  fields}
In this section we will study
the non commutative quantum harmonic
oscillator and how it can be used to define a non commutative field
theory. As a warm up exercise let us consider first the standard
bidimensional quantum harmonic oscillator described by the following
Hamiltonian operator
\begin{equation}
H=\frac{\omega}2[p_1^2 +p_2^2+q_1^2+q_2^2], \label{ha}
\end{equation}
with standard commutation relations ($i,j=1,2$)
\begin{eqnarray}
\left[q_i,q_j\right]&=&\left[p_i,p_j \right]=0, \nonumber
\\
\left[q_i,p_j\right] &=& i \delta_{ij}. \label{us}
\end{eqnarray}
Note that variables $\{q_i,p_j\}$ are dimensionless and are related
with usual ones $\{Q_i,P_j\}$ by $q_i=\sqrt{m\omega}Q_i$ and
$p_i=(m\omega)^{-1/2} P_i$.

The dynamics of this system is described by the Heisenberg equations
\begin{equation}
i\,\frac {d{\cal O}}{dt}=[{\cal O}, H]
\end{equation}
which, specified to (\ref{ha}) and (\ref{us}) give rise to
\begin{subequations}
\label{mov}
\begin{eqnarray}
\dot{q_i}&=&\omega p_i, \label{eq1}
\\
\dot{p_i}&=& -\omega q_i. \label{eq2}
\end{eqnarray}
\end{subequations}
\noindent The system (\ref{mov}) is equivalent to the very well known
second order differential equation
\begin{equation}
\ddot{q_i}= - \omega^2 \,q_i. \label{sec}
\end{equation}
Therefore, the solution of (\ref{mov}) turn out to be
\begin{eqnarray}
 \label{solu2}
q_i (t) &=& A_i ~ e^{i\omega t} + B_i ~e^{-i \omega t},\nonumber
\\
p_i(t) &=& i A_i ~e^{i \omega t} - i ~ B_i ~e^{-i \omega t}.
\end{eqnarray}

The algebra of operators $A_i$ and $B_i$ can be fixed using the
canonical algebra (\ref{us}). Indeed, replacing (\ref{solu2}) in
(\ref{us}) we find that
\begin{eqnarray}
\left[A_i, A_j \right] &=&\left[ B_i, B_j \right] = 0, \nonumber
\\
\left[ A_i,B_j \right] &=& -\frac{1}{2}\delta_{ij}. \label{algeb}
\end{eqnarray}
Then we can identify
\begin{equation}
\sqrt{2}A_i \rightarrow a^{\dagger}_i, \,\,\,\,\,\,\,\,\,\,\,\,\,\,\,
\sqrt{2}B_i \rightarrow a_i,
\label{sealge}
\end{equation}
and  the algebra (\ref{algeb}) becomes
\begin{eqnarray}
\left[a_i, a_j \right] &=&\left[ a^{\dagger}_i, a^{\dagger}_j \right]
            = 0, \nonumber
\\
\left[ a_i,a^{\dagger}_j \right] &=& \delta_{ij} . \label{algeb1}
\end{eqnarray}

In terms of $a_i^{\dagger}$ and $a_i$, as is well known, one find that
the Hamiltonian becomes
\begin{equation}
H = \omega ( a^{\dagger}_1 a_1+ a^{\dagger}_2 a_2 + 1), \label{haa}
\end{equation}
The construction of the Hilbert space is straightforward.

Using these results let us solve the noncommutative harmonic
oscillator described by the Hamiltonian (\ref{ha}) but commutation
relations deformed as follows:
\begin{equation}
\label{conmus}
[q_i,q_j]=i\theta \epsilon_{ij},\,\,\,\,\,\,\,[p_i,p_j]=i{\cal
B}, \epsilon_{ij}\,\,\,\,\,\,\,[q_i,p_j]=i\delta _{ij},
\end{equation}
where $\theta$ and ${\cal B}$ are parameters \lq \lq measuring"
noncommutativity in $q$ and $p$ respectively\footnote{Note that this
  parameters are dimensionless in our notation, but they actually have
dimensions in the standard variables.}.

Using the Hamiltonian (\ref{ha}) and (\ref{conmus}) one find that
equations of motion are
\begin{subequations}
\label{ncq}
\begin{eqnarray}
\dot{q_i}&=&\omega (p_i +\theta\,\epsilon_{ij} q_j),
\\
\dot{p_i}&=&\omega (-q_i +{\cal B}\,\epsilon_{ij} p_j). \label{la}
\end{eqnarray}
\end{subequations}

Following the same procedure previously discussed, one find that the
analogous to (\ref{sec}) turn out to be now
\begin{equation}
\ddot{q_i}=\omega(\theta + {\cal
  B})\,\epsilon_{ij}\,\dot{q_j}+\omega^2({\cal B}\theta
-1)\,q_i. \label{senc}
\end{equation}
Although this set of equations are coupled ones, one decouple these
equations using the following trick: let us define the complex
variable $Z= q_1 + i q_2 $, then (\ref{senc}) can be written as
\begin{equation}
\ddot{Z}=-i\,\omega(\theta + {\cal B})\,\dot{Z}+\omega^2({\cal
  B}\theta -1)\,Z, \label{senc1}
\end{equation}
and a similar equation for the conjugate $Z^\dag$.
This problem is formally is equivalent to the damped harmonic oscillator.

As in the standard case, we look now for a solution with the shape
   \[
   Z(t) = e^{ \alpha t},
   \]
which imply that
\[
\alpha^2 + i \omega(\theta + {\cal B})\,\alpha - \omega^2({\cal
  B}\theta -1)=0 ,
\]
and as a consequence the possible values of $\alpha$ are
\begin{equation}
\label{solal}
\frac{\alpha_\pm}{\omega}=i\,\left(-\frac{\theta +{\cal B}}{2} \pm
\sqrt{1+\left(\frac{\theta -{\cal B}}{2}\right)^2}\right).
\end{equation}

Thus, the general solution of the noncommutative harmonic oscillator is
\begin{equation}
\label{z}
Z(t)=A_+\,e^{i\alpha_+ t} + A_-\,e^{i\alpha_- t},
\end{equation}
where $A_\pm$ are complex operators. Note that there is a redefinition
of $\alpha_\pm$ since we have factorized the $i$ in (\ref{solal}). It
is interesting to note also that the solution (\ref{z}) is a
superposition of two oscillation modes, one positive (because
$\alpha_+>0$ ) and other negative (because
$\alpha_-<0$). Therefore, the solution has the same structure of the
standard case, but there is an asymmetry due to the fact that
$|\alpha_-|\neq \alpha_+$.

Of course from (\ref{z}) it is possible to compute $q_i (t)$, and from
the equation of motion we obtain $p_i$
\begin{subequations}
\label{qa}
\begin{eqnarray}
q_j&=&\frac{(-i)^{j-1}}{2}\bigg[ a+b-(-)^{j}(a^\dag + b^\dag)
\bigg],\\
p_j&=&-\frac{(-i)^{j}}{2}\bigg[\lambda_+
  (a+(-)^{j}a^\dag)+\lambda_-(b+(-)^{j}b^\dag) \bigg].
\end{eqnarray}
\end{subequations}
with $a=A_+e^{i\alpha_+t}$, $b=A_-e^{i\alpha_-t}$ and
$\lambda_\pm=\theta+ \alpha_\pm/\omega$

Following the example of the commutative case, we must find the
commutation relation between operators $A_\pm$ and $A_{\pm}^\dag$ from
the known relations (\ref{conmus}).

Since the result of this commutators do not depend on time, the
following condition fulfills
$$
[A_+,A_-]=0=[A_+,A_{-}^\dag].
$$
Note that the remaining commutators, $[A_\pm,A_{\pm}^\dag]$ are
obtained from the conditions $[q_1,q_2]=i\theta$ and $[p_1,p_2]=iB$
and then, the condition $[q_1,p_1]=i$ is just a consistency
check. After a straightforward calculation we obtain
\begin{equation}
\label{aes}
{[}A_\pm,A_{\pm}^\dag{]}=\mp 2 \frac{1+\theta\lambda_\mp}
{\lambda_+ -\lambda_-}.
\end{equation}

This equation shows that non commutative harmonic oscillator in two
dimensions is equivalent to two one-dimensional harmonic
oscillator.

In fact, by a rescaling of $A_\pm$ operators
$$
\tilde{A}_+=\left(\frac{\lambda_+ -\lambda_-}
{2(1+\theta\lambda_-)}\right)^{1/2}~A_+^\dag,~~~~~~~~
\tilde{A}_-=\left(\frac{\lambda_+ -\lambda_-}
{2(1+\theta\lambda_+)}\right)^{1/2}~A_-,
$$
and similar relations for $\tilde{A\pm}^\dag$, we find
$$
[\tilde{A}_+,\tilde{A}_+^\dag]=1=[\tilde{A}_-,\tilde{A}_-^\dag].
$$
Operators $\tilde{A}_\pm$ play the role of standard lowering
and rising operator as in the commutative case. The Hamiltonian of
this two dimensional non commutative harmonic oscillator is
\begin{equation}
\label{hamil}
H=\omega_+(\tilde{A}_+\tilde{A}_+^\dag+1/2)+\omega_-(\tilde{A}_-
\tilde{A}_-^\dag+1/2).
\end{equation}
with $\omega_\pm=\sqrt{1+\left(\frac{\theta - B}{2}\right)^2}\pm
\left(\frac{\theta+B}{2}\right)$.

Previous result is the starting point for constructing noncommutative
complex scalar field theory \cite{nos2a}.  Consider
the standard relativistic Hamiltonian density for a complex scalar
field
\begin{equation}
\label{hamrel}
{\cal H}=\Pi^\dag\Pi+{\boldsymbol {\nabla}}\Phi^\dag
{\boldsymbol{ \nabla}} \Phi +m^2\Phi^\dag\Phi
\end{equation}
plus non standard commutation relations
\begin{subequations}
\label{defcons}
\begin{eqnarray}
{[}\Phi(\x),\Phi^\dag({\x}'){]}&=&\theta\delta^3(\x-{\x}'),
\\
{[}\Pi(\x),\Pi^\dag({\x}'){]}&=&B\delta^3(\x-{\x}'),
\\
{[}\Phi(\x),\Pi({\x}'){]}&=&\delta^3(\x-{\x}'),
\end{eqnarray}
\end{subequations}
where $\theta$ and $B$ parameterizes the non commutativity in the
field space and have dimensions of energy$^{-1}$ and energy,
respectively.

In the standard case $(\theta=0=B)$, quantized fields are a
superposition of quantum harmonic oscillators with frequency
$\omega(\p)=\sqrt{\p^2+m^2}$, one for each value of
momenta $\p$. The structure of these linear superpositions are
given by (\ref{solu2}).

For non commutative fields, the constructions proceeds in a
similar way.  The analog of (\ref{solu2}) is given by (\ref{qa}) once
they are expressed in terms of operators $\tilde{A}_\pm$.
Therefore we consider now a linear superposition with
\begin{subequations}
\label{ncfields}
\begin{eqnarray}
\Phi(\x)&=&\int \frac{d^3p}{(2\pi)^3}\frac{1}{\sqrt{\omega}}\left[
\eta\epsilon_ 1 a_\p~e^{i\p\x} + \epsilon_2
b_\p^\dag~e^{-i\p\x}\right],
\\
\Pi(\x)&=&i\int \frac{d^3p}{(2\pi)^3}\sqrt{\omega}\left[
-\epsilon_ 1 a_\p~e^{i\p\x} +\eta \epsilon_2
b_\p^\dag~e^{-i\p\x}\right],
\end{eqnarray}
\end{subequations}
where coefficients $\eta$ and $\epsilon_i$ are those appearing in
the previously mentioned linear superposition, but with $\omega$
replaced by $\omega(\p)$. We are following  notation of
\cite{nos2a} where $\eta=\lambda_+$ and
$\epsilon_1^2=(\lambda_++B)(\lambda_+^2+1)^{-1}$ and
$\epsilon_1^2=(\lambda_+-\theta)(\lambda_+^2+1)^{-1}$. Note that
dependences on ${\bf p}$ are inherited from $\omega(\p)$.

Operators $a,b$, by other hand,  satisfy the canonical algebra
$$
[a_\p,a_{{\p}'}^\dag] =(2\pi)^3\delta^3(\p-\p'),
\,\,\,\,\,\,\,\,\,
[b_\p,b_{{\p}'}^\dag] =(2\pi)^3\delta^3(\p-\p'),
$$
and are in correspondence with $\tilde{A}_\pm$. Is straightforward
to prove that fields constructed in this way satisfy the
commutation relations (\ref{defcons}).

The Hamiltonian of this theory
$$
H=\int d^3x{\cal H}(x),
$$
with the density (\ref{hamrel}) expressed in terms of the non
commutative fields (\ref{ncfields}) is a superposition of two
anisotropic oscillators (\ref{hamil}) where frequencies are now
$\omega(\p)$. That is
\begin{eqnarray}
\label{nchamsf}
H&=&\int
\frac{d^3p}{(2\pi)^3}\left[E_+(\p)\left(a_{{\p}}^\dag
a_\p+\frac{1}{2}\right)+E_-(\p)\left(b_{{\p}}^\dag
b_\p+\frac{1}{2}\right)\right],
\end{eqnarray}
where energies are
\begin{eqnarray}
\label{scalener}
E_\pm&=&\sqrt{\omega^2(\p)+\frac{1}{4}\bigg[B-\theta \omega^ 2
(\p)\bigg]^2}\pm\frac{1}2\left(B+\theta\omega^2(\p)\right ) .
\end{eqnarray}

This shows that the free non commutative complex scalar fields is a
theory with two types of particles with different dispersion
relation. This asymmetry can be interpreted as a particle-antiparticle
asymmetry and their phenomenological consequences were
explored in \cite{nos2,nos5}.

A natural question raised by this approach is what happens with other
fields as gauge and fermionic fields. Let us discuss gauge fields in next subsection and
postpone fermionic fields and its phenomenology to subsequent sections.

\subsection{Non commutative Gauge Fields}
We start the discussion with the $U(1)$ gauge field. As in the
previous case, the theory is defined by the standard Hamiltonian \cite{gl} (for a previous approach see also\cite{jackiw}).

\begin{equation}
\label{5}
H = \int d^3x \left( \frac{1}{2}{\vec \pi}^2 + \frac{1}{2}{\vec
B}^2+A_0\,\vec\nabla\cdot\vec\pi\right),
\end{equation}
plus a set of deformed Poisson bracket structure
\begin{subequations}
\label{gaupb}
\begin{eqnarray}
{[} A_i ({\vec x}),A_j ({\vec y}){]}_{PB} &=&0, \nonumber
\\
{[} A_i ({\vec x}),\pi_j ({\vec y}){]}_{PB} &=& \delta_{ij}
\delta ({\vec x}-{\vec y}), \label{}
\\
{[}\pi_i ({\vec x}),\pi_j ({\vec y}){]}_{PB}&=&\theta_{ij}\delta
({\vec x}-{\vec y}), \nonumber
\end{eqnarray}
\end{subequations}
where $\theta_{ij}$ is the most general antisymmetric three
dimensional matrix $\theta_{ij}=\epsilon_{ijk}\theta_k$. Note that
this term modifies the infrared sector of the theory due to the
dimension of $\theta_k$.

Poisson brackets (\ref{gaupb}) break Lorentz invariance and also the
gauge symmetry (GS). In fact, gauge transformation is generated by
the Gauss law, which is just the condition that primary constrain
preserves in time. The primary constrain is not modified, but its
time preservation indeed changes because of extra terms coming from
the modified Poisson bracket structure.

In order to study just LIV, preserving gauge symmetry, a modified
Gauss law must be introduced. A direct calculation shows that
\begin{equation}
\label{ngl}
\chi=\bn\cdot{\boldsymbol \pi} -\bth\cdot\B,
\end{equation}
with $\bth\cdot\B=\theta_i\epsilon_{ijk}\partial_jA_k$ is the
modified generator for the gauge symmetry. That is, if $\alpha(\x)$
is an arbitrary and real function, then
\begin{subequations}
\label{gf}
\begin{eqnarray}
\delta A_i(\x)&=&{[}A_i(\x),\Delta_\alpha{]}_{PB}
\nonumber
\\
&=&\partial_i\alpha(\x),
\\
\delta \pi_i(\x)&=&{[}\pi_i(\x),\Delta_\alpha{]}_{PB}
\nonumber
\\
&=&0.
\end{eqnarray}
\end{subequations}
where the gauge transform operator $\Delta_\alpha$ is
\begin{equation}
 \Delta_\alpha=\int d^3y~\alpha({\bf y })\bn[ {\boldsymbol \pi}
({\bf y}) + \bth\times {\bf A}({\bf y})].
\end{equation}

This last relation allows to write the modified Hamiltonian which
includes now this new gauge symmetry generator, namely
\begin{equation}
\label{modham}
H=\int d^3x \left(\frac{1}{2}{\boldsymbol \pi}^2 + \frac{1}{2}
\B^2 +A_0 \bn \bigg[{\boldsymbol \pi} +
\bth\times {\bf A}\bigg]\right).
\end{equation}

The Hamiltonian (\ref{modham}) with the Poisson bracket structure
(\ref{gaupb}) defines a $U(1)$ gauge field theory which breaks
Lorentz symmetry.

This model originates modified Maxwell equations
\begin{eqnarray}
\dot{A}_i&=&\pi_i - \partial_iA_0,
\\
\dot{\pi}_i&=&({\boldsymbol \pi}\times\bth)_i -(\bn\times\B)_i.
\end {eqnarray}
First equation is the electric field definition
$$
-E_i\equiv \dot{A}_i+\partial_i A_0,
$$
which allow to write the remaining equations in the usual form.
Including the modified Gauss law they read
\begin{subequations}
\label{mxw}
\begin{eqnarray}
\label{mxw1}
\bn\cdot{\bf E}&=&-{\bth}\cdot\B,
\\
\label{mxw2}
\bn\cdot\B&=&0,
\\
\label{mxw3}
\bn\times{\bf E}+\frac{\partial\B}{\partial t}&=&0.
\\
\label{mxw4}
\bn\times\B -\frac{\partial {\bf E}}{\partial t} &=&
\bth\times {\bf E}.
\end{eqnarray}
\end{subequations}

In this equation $\theta$ plays the role of a \lq\lq source \rq\rq~
term that can be interpreted as a polarization charge and induced
current in a medium, in a similar way to the standard electromagnetic
theory. In section IV  we will discuss the physical
implications of this fact.

Now we would like to discuss other issue, related with the Lagrangian
formulation of our approach. From here, the generalization for other
gauge groups will be straightforward \cite{nos3}.

The set of equations (\ref{mxw}) can be obtained from a Lagrangian
which is constructed from the Hamiltonian (\ref{modham}) in the
standard way only if  dynamical variables are canonical. Then, we
need to find a transformation from variables $\{\pi_i,A_j\}$ to
variables $\{\tilde{\pi}_i,\tilde{A}_j\}$ such that the Poison bracket
structure (\ref{gaupb}) maps to the canonical one.

This procedure is completely analog to the change of variables,  in
previous section, which takes the non commutative
phase space variables to a set of standard rising and lowering
operators. In the present case, transformations read
\begin{equation}
\label{ptop}
\tilde{\pi}_i=\pi_i+\frac{1}{2}(\bth\times{\bf A})_i,\,
\,\,\,\,\,\,\,\,\,\,\,\,\,\,\,
\tilde{ A}_i=A_i.
\end{equation}
and the Lagrangian turn out to be
\begin{eqnarray}
\label{lagmxw}
\label{lag}
L &=& \int d^3 x ~(\tilde{\pi}_i {\dot A}_i - H), \nonumber
\\
&=& \int d^3x \left( {\bf E}^2 - {\B}^2 + \frac{1}{2}A_0
{\bth}\cdot{\B} - \frac{1}{2}{\bf A}\cdot{\bth} \times
{\bf E}) \right).
\end{eqnarray}

Using the standard definition for $F_{\mu\nu}$ and ${\tilde F}^{\mu
\nu}= \frac{1}{2}\epsilon^{\mu\nu\lambda\rho}F_{\lambda\rho}$, ($\mu,\nu,..=0,1,...3$) one
finds that
\begin{equation}
L = \int \left( -\frac{1}{4} F_{\mu \nu} F^{\mu \nu} + \frac{1}{2}
\theta_\mu {\tilde F}^{\mu \nu}A_\nu\right)d^3x. \label{21}
\end{equation}

This approach can be generalized to others gauge groups. For
instance  in \cite{nos3}, the $SU(2)$ gauge group was studied and it
was shown that the Lagrangian density that generalizes (\ref{21}) is
\begin{equation}
L = - \frac{1}{2} {\mbox{tr}} \left\{F_{\mu \nu}
F^{\mu\nu}\right\} + 2\,\theta^\mu \epsilon_{\mu \nu \rho \sigma}
\mbox{tr}\left(A^\nu F^{\rho \sigma} + \frac{2}{3}g A^\nu A^\rho
A^\sigma \right), \label{lag2}
\end{equation}

Finally, let us point out that the Chern-Simons term appearing in
this formulation is not a perturbative contribution, it
appears indeed at the same footing as $F^2$ in the $g$ expansion.

We would like now to call the attention on the fact that here we have
considered commutator deformations only in the momenta, although we
start the discussion with the complex non commutative scalar field
where deformations in fields commutators also appears.

It is natural to ask, therefore, what kind of modifications suffers
Maxwell theory if we consider a Poisson bracket structure modified as
\begin{equation}
\label{uvmaxmod}
[A_i(x),A_j(x)]=\epsilon_{ijk}\theta_k\delta(x-y),
\end{equation}
with $\theta_k$ a Lorentz violating vector which plays a role of
ultraviolet energy scale. The rest Poisson brackets are canonical.

This theory was considered in \cite{glp} and we will not give details
on its construction. We would like just to say that, as cases
presented before, is possible to restore the gauge symmetry.

The main phenomenological feature of this approach is that it
presents a birefringence effect with polarization planes shifted by
an amount proportional to
$$
\omega^2\theta\cos \alpha,
$$
where $\theta$ is the modulus of $\theta_k$ and $\alpha$, the angle
between the wave vector $\vec{k}$ and $\theta_k$. This frequency
dependent behavior is distinctive and it does not appear when
noncommutativity in space-time is considered. It is a pure field
theory result.

Our last example of non commutative fields is the fermionic case. The
next section is devoted to this issue.

\section{Neutrino physics and noncommutative fermionic fields}

A natural question, considering all the previous discussion, is what
happens with the fermions fields under a deformation of Poisson
structure.

From the phenomenological point of view, our approach is a
mechanism that offers an alternative way out to the problem of
neutrino oscillations. Let us emphasize that several papers have
already dealt with effects of Lorentz \cite{8} and CPT violation
\cite{9} in this scenario, however the model we present here has the
advantage that depends on a few parameters, as was discussed in
previous sections.

Since we are mainly interested in the neutrino sector, and
particularly in the problem of oscillations, let us briefly review
the situation \cite{neutrino}. Neutrino oscillation is a
phenomenological model proposed to explain the deficit of solar and
atmospheric neutrinos in fluxes measured on earth \cite{more}.

The key idea of this mechanism is to assume that neutrino are massive
particles which, upon propagation, oscillates between different
flavor eigenstates. In its simplest form, the oscillation between two
flavors $i,j$ is considered and it is shown that the probability
$P_{i\to j}(t)$ for specie $i$ to oscillates to $j$ after a time $t$
is
\begin{equation}
\label{Pij}
P_{i\to j}(t)=\sin^2(2\theta_{ij})\sin^2\left(\frac{(E_i - E_j)t}{2}
\right),
\end{equation}
where $\theta_{ij}$ is the mixing angle\footnote{This angle is
introduced to take into account the fact that what propagates is a
linear superposition of mass eigenstates.} and $E_{i(j)}$ is the
energy of $i(j)$ species.

It is interesting to note that non vanishing oscillation probability
occurs in free space only if there is a non zero $\Delta
E_{ij}=E_i-E_j$ and
$\theta_{ij}\neq 0$. Then, if neutrino have equal masses or they are
zero, oscillation does not come out.

The standard scenario assumes that neutrino species have small masses
and therefore, the probability for oscillation of two neutrino
$\nu_i$, $\nu_j$, in traversing a path length $L$ turns out to be
\begin{equation}
\label{standprob}
P_{\nu_i \to \nu_j}(L) = \sin^2(2\theta_{ij})
\sin^2\left( \frac{1.27\Delta m^2_{ij} L}{E} \right),
\end{equation}
where $\Delta m^2_{ij} =m^2_i -m^2_j$ is taken in (eV)$^2$, the
neutrino energy $E$ in MeV and $L$ in meters.

Clearly, with three families of neutrino there can be only two
independent combinations of squared mass differences, lets say
$\Delta m^2_{12},\Delta m^2_{23}$ from which a solution for solar
neutrino as well as atmospheric neutrino puzzles is found. The
bounds for this masses in this scenario are \cite{neutrino}
\begin{equation}
\label{masses}
\Delta m^2_{12}\leq 10^{-4}\mbox{eV}^2,\,\,\,\,\,\,\,\,\,\,
10^{-3}\leq \mbox{eV}^2 \Delta m^2_{23}\leq 10^{-2}\mbox{eV}^2.
\end{equation}
From here, the bound for $\Delta m^2_{13}$ is fixed.

LSND (Liquid Scintillator Neutrino Detector at Los Alamos)
\cite{lsnd} is one of the several experiments that have looked for
neutrino oscillations. It has used muon sources from the decay
$\pi^+\to \mu^+ +\nu_\mu$. These muons decay through $\mu^+\to
e^++\nu_e+\bar{\nu}_\mu$ and, after a $30$ meters long path, the
experiment finds the oscillation channel $\bar{\nu}_\mu\to
\bar{\nu}_e$ at $20~\mbox{MeV}\leq E_{\nu_\mu}\leq 58.2~\mbox{MeV}
$ with probability of $0.26\%$. According to (\ref{standprob}),  the
mass difference involved in this process should be
\begin{equation}
\label{mlsnd}
\Delta m^2<1~\mbox{eV}.
\end{equation}

 Analysis of the MiniBooNE experiment \cite{mb} will confirm or
discard this result. However, if it is true, we must face a puzzle
within the context of three families neutrinos. An explanation of
this anomaly, compatible with the standard model, may require the
existence of sterile neutrinos \cite{steril}.

There is a different approach that might solve the puzzle. As was
pointed out first by Coleman and Glashow \cite{colgla} and
more extensively developed by Kostelecky and collaborators
\cite{kostel}, a departure from Lorentz and/or CPT symmetry in the
neutrino sector, gives raise to oscillations as neutrino propagates
in free space, even for massless neutrinos.

From (\ref{Pij}) is clear  that the oscillation probability does not
vanish if the mixing angle is not zero and the energy of specie $1$
is different from specie $2$. This condition satisfies if neutrino
masses are different in a theory respecting Lorentz and CPT symmetry
(the standard case), but if one of these  symmetries is broken, a non
zero probability can be obtained.

Note that it is enough to have different dispersion relations for
different neutrino species in order to fulfill previous requirements.
But this is indeed the case for noncommutative fields approach if we
choose non standard commutators among different particle species. In
fact, noncommutative scalar field is the first example of this kind,
since there field $\phi$ and its conjugate $\phi^\dag$ have a non
standard Poisson bracket.

The final result is that, due to non standard anticommutators for
massless fermionic fields, dispersion relations are species
dependent and a non vanishing oscillation probability is obtained. In
what follows we will discuss the technicalities of this approach
\cite{last}.

 In the chiral basis, which is more convenient for
our analysis, the Hamiltonian density has the form
\begin{equation}
{\cal H} = i\left( \psi_L^{i\dagger} {\vec \sigma} \cdot~ \vec{\nabla}
  \psi^{i}_L -
\psi_R^{i\dagger} {\vec \sigma} \cdot~ \vec{\nabla} \psi^{i}_R
\right), \label{dh1}
\end{equation}
where the superscript $i=\{1,2\}$ runs over the flavor quantum number
(sum over repeated indexes).

The non-commutative theory is obtained by
deforming the canonical anti-commutation relations while maintaining
the form of the Hamiltonian density \eqref{dh1}. We postulate the
deformed equal-time anti-commutation relations to have the form (with all others
vanishing)
\begin{eqnarray}
\{ \psi^i_L({\bf x}),\psi^{j\dagger}_L ({\bf y})\} &=& A^{ij}
~\delta^{(3)} ({\bf x} - {\bf y}),
\label{ant1}
\\
\{ \psi^i_R ({\bf x}),\psi^{j\dagger}_R({\bf y})\} &=& B^{ij}
~\delta^{(3)} ({\bf x} - {\bf y}),
\label{ant2}
\end{eqnarray}
where $A^{ij}$ and $B^{ij}$ are $2\times 2$ matrices with constant,
complex elements in general, but if we want to maintain rotational
invariance, they can be chosen to have the forms
\begin{equation}
\label{abmatri}
A = \left(\begin{array}{cc} 1 & \alpha \\
\alpha^\ast & 1 \end{array}\right),~~~~~~~~~
B = \left(\begin{array}{cc} 1 & \beta \\
\beta^\ast & 1 \end{array}\right),
\end{equation}
so that the complex parameters $\alpha,\beta$ can be thought of as the
parameters of deformation. Clearly, these deformed anti-commutation
relations reduce to the conventional ones when the parameters of
deformation vanish.

Equation of motions in the momentum space read
\begin{eqnarray}
E\psi^i_L &=& -A^{ij} \left( {\vec \sigma}\cdot {\vec p}\
\psi^j_L\right),
\label{mom1}
\\
E\psi^i_R&=& B^{ij}\left( {\vec \sigma}\cdot {\vec p}\
\psi^j_R\right). \label{mom2}
\end{eqnarray}

The energy spectrum can be find independently for \eqref{mom1} and
\eqref{mom2}. If we take the first one, it is straightforward to find
a diagonalization matrix for $A$
\begin{equation}
D= \frac{1}{\sqrt{2}} \left(\begin{array}{cc} \frac{|\alpha|}{\alpha}
  & 1 \\
-\frac{|\alpha|}{\alpha} &1\end{array}\right),\
\end{equation}
and then
\begin{eqnarray}
E^{1}_\pm &=& \pm \left( 1 +|\alpha| \right) |{\vec p}|,
\nonumber
\\
E^{2}_\pm &=& \pm \left( 1 -|\alpha| \right) |{\vec
  p}|,
 \label{disp2}
\end{eqnarray}
are the eigenvalues in \eqref{mom1}.

A similar analysis for \eqref{mom2} gives
\begin{eqnarray}
E^{1}_\pm &=& \pm \left( 1 +|\beta| \right) |{\vec p}|,
\nonumber
\\
E^{2}_\pm &=& \pm \left( 1 -|\beta| \right)
|{\vec p}|.
 \label{disp22}
 \end{eqnarray}

Since $\psi^i_{L}$ does not diagonalize the Hamiltonian --eigenvectors
are $D\psi^i_L$, which are a linear combination of $\psi^1_L$ and
$\psi^2_L$-- the time evolution of this field gives rise to a linear
combination of species 1 and 2, namely
  a neutrino initially in the state $\psi_{L}^{1}$ would evolve in
time as
\begin{eqnarray}
\psi^1_L (t) &=& \cos \theta_{12} ~{\tilde \psi}^1_L (t) -
\sin \theta_{12} ~{\tilde \psi}^2_L (t) \nonumber
\\
&=& \biggl[ \left( \cos^2 \theta_{12}~e^{-iE_+^{1} t} +
  \sin ^2\theta_{12}~ e^{-iE_+^{2} t}\right) \psi^1_L(0)
\nonumber
\\
&+& \frac{1}{2} \sin2 \theta_{12} \left( e^{-iE_+^{1} t} -
e^{-iE_+^{2}
    t}\right) \psi_L^2 (0)\biggr]e^{i{\vec p}\cdot{\vec x}}.  \nonumber
\end{eqnarray}
Therefore, after a path of length $L$, the probability of finding the
state $\psi^2_L$ in the beam is given by
\begin{equation}
P_{\nu_1 \rightarrow \nu_2}= \sin^2 \left(2\theta_{12} \right) ~
\sin^2
\left( \vert \alpha \vert~E ~L\right),\label{prob3}
\end{equation}

\noindent where we have used the fact that for $|\alpha|\ll 1,
E\approx |\vec{p}|$.

Same arguments demonstrate that, for antineutrino propagation, the
probability for oscillations is
\begin{equation}
P_{{{\bar \nu}_1 \rightarrow {\bar \nu}_2}}= \sin^2
\left(2\theta_{12}
\right) ~
\sin^2 \left( \vert \beta \vert~E~ L\right). \label{prob4}
\end{equation}
The important thing to note here is that
\begin{equation}
P_{\nu_1 \rightarrow \nu_2} \neq P_{{\bar \nu}_1 \rightarrow
{\bar \nu}_2 }, \label{000}
\end{equation}
which is a consequence of $CPT$ and Lorentz invariance violation.

Two comments are  in order here. First, noncommutative fermionic fields
have a mixing angle equals $\pi/4$, and then is consistent with the so
called Large Mixing Angle (LMA) scenario. However, we have introduced
in all the discussion a $\theta_{ij}$, in order to mimic that
standard case, but it does not appear in a pure noncommutative
fermionic theory. Secondly, our description has a linear dependence
on the energy, what could be disturbing, but is a natural  consequence of the
fact that we have incorporated two dimensionless parameters on the dispersion relation.

From previous results we can compute bounds for deformation
parameters as follow. Assuming that flavor oscillations involve only
 pairs of neutrinos, when dealing with three families we must
generalize the parameter $\alpha$ to $\alpha_{ij}$ (and the same for
antineutrinos). Therefore, flavor oscillation probabilities are
\begin{eqnarray}
P_{\nu_{i}\rightarrow \nu_{j}}&=& \sin^{2}
\bigg(2\theta_{ij}\bigg) \sin^{2}
\bigg(|\alpha_{ij}| EL\bigg),\label{probij}
\\
P_{\bar{\nu}_{i}\rightarrow \bar{\nu}_{j}} &=& \sin^{2}
\bigg(2\theta_{ij}\bigg) \sin^{2}
\bigg(|\beta_{ij}| EL\bigg).\label{probijbar}
\end{eqnarray}

From the solar neutrino experiments, oscillations of the
flavors $1\rightarrow 2$ are involved, while the channel
$2\rightarrow 3$ is related with atmospheric neutrino oscillations.
Therefore, from \eqref{probij} next bounds on $\alpha$ parameter is
found
\begin{eqnarray}
|\alpha_{12}| &\leq& 10^{-17}.\label{result1}
\\
|\alpha_{23}| &\leq& 10^{-22}.\label{result2}
\end{eqnarray}

LSND, by other hand, due to the fact that involves antineutrinos,
gives a bound for $\beta$ parameter
\begin{equation}
|\beta_{12}|\leq 10^{-16}.\label{result3}
\end{equation}
It is clear that within this scenario, all the experimental results
can be naturally explained.

Finally we would like to call the attention on the fact that here
neutrino are massless particles and, looking at dispersion relations,
we realize that the origin of the energy difference between species
can be understood also as a difference in the propagation velocities.
Since particles are massless, we would say that particles propagates
with different speed of light.

\section{Phenomenological consequences of noncommutative fields in
the early universe.}

This section is devoted to analyze possible phenomenological
consequences derived from models based on noncommutative fields.

A consequences of this model which is common for scalars and massless
fermions is the asymmetry of the dispersion relation for
particles and antiparticles.

For the scalar field, situation is rather simple and the asymmetry
can be checked at the level of quantum Hamiltonian
\eqref{nchamsf} or at the level dispersion relation
\eqref{scalener}.  This theory has two scales, one infrared $B$ and other ultraviolet
$\theta$. At momentum ${\bf p}$ such that
$B<<\omega({\bf p})<<\theta^{-1}$ energy \eqref{scalener} satisfy
$ E_+\sim E_-\sim \omega({\bf p})$ and we are in the Lorentz invariant
region.

Consider a system with this two types of particles in thermodynamical
equilibrium at temperature $T$. The density $n/V$ of each specie
contained in a volume $V$ (with zero chemical potential) is
\cite{nos5}
\begin{eqnarray}
\label{nmas}
n_+=4\pi\int_0^\infty\frac{p^2dp}{e^{{E_+}/T}-1},
\\
\label{nmenos}
n_-=4\pi\int_0^\infty\frac{p^2dp}{e^{{E_-}/T}-1}.
\end{eqnarray}

For a temperature $T$ such that $\theta T<<B/T<<m/T<<1$, there is a
tiny asymmetry in the dispersion relation due to the infrared scale
and then, a tiny asymmetry in the content of baryonic
matter-antimatter content in the volume $V$. In fact
\begin{equation}
\label{barasymm}
\frac{n_+ - n_-}{n_-}\sim \frac{B}{T},
\end{equation}
as it is expected if CPT violating effects are tiny.

This example shows that a baryon asymmetry can be generated without
departure from thermal equilibrium and it suggests a critical
reevaluation of the third criterion of Sakharov for baryogenesis
\cite{sakha}.

Since this effect is related only with the asymmetry on the
dispersion relation for particles and antiparticles, one could wonder
what happens with fermions.

Situation is similar to the previous one. There is an asymmetry
due to the different dispersion relations for neutrinos and
antineutrinos\footnote{In this case, however, the distinction
between particles and antiparticles is more subtle, but the
quantum Hamiltonian of the theory can also be written as two types
of particles with different frequencies\cite{lastt}} and therefore
a the ratio of the neutrino density to the antineutrino density,
in equilibrium at certain temperature is different from one.

Calculation in this case is more involved due to the presence of
different flavors and to the oscillations between them. For two
species, however, this problem is formally equivalent to a quantum
mechanical two level system with a Hamiltonian which is responsible
for the for inducing transition between levels.

The crucial step is to identify this Hamiltonian responsible for the
transitions, lets say $\nu_e\to\nu_\mu$. According to Stodolsky
\cite{sto} and others \cite{others} this Hamiltonian is
\begin{equation}
\label{stodol}
H=\vec{\sigma}\cdot\vec{V},
\end{equation}
with $|\vec{V}| =|E_1-E_2|$.

Following the results of previous section, we find that the
Hamiltonian for a system violating CPT and Lorentz symmetries can be
written in this two dimensional space as
\begin{eqnarray}
\label{hmas}
H_+=2|\alpha|\vec{\sigma}\cdot\vec{p},
\\
H_-=2|\beta|\vec{\sigma}\cdot\vec{p}.
\end{eqnarray}

Then, if there is a CPT and Lorentz invariance violation,
oscillations neutrino-neutrino and antineutrino-antineutrino take
place with different probabilities leading to a neutrino asymmetry.

With present data on neutrino experiments, this asymmetry could be
evaluated, however, it seems to premature to carry this analysis at
this stage, when still LSND experiment requires a confirmation.

We will finish this section with some comments on the
cosmological implications of noncommutative $U(1)$ gauge field. As
we mentioned in section II.A, modified Maxwell equations
\eqref{mxw} mimics standard Maxwell equations with sources, but there
is an important difference:electrostatic and magnetostatic appears
mixed and then the presence of polarization implies a magnetization
and viceversa.

Remarkably, is this structure which offers an alternative to the
dynamo mechanism to generate the so called {\it primordial magnetic
field}. In fact, it was shown in \cite{nos3} that modified equations
admits a solution with the shape
\begin{eqnarray}
{\bf B}&=&\B^{(0)} +\B^{(2)}+\B^{(4)}+\ldots\B^{(2n)}+\ldots,
\\
{\bf E}&=&{\bf E}^{(0)} +{\bf
E}^{(1)} + {\bf E}^{(3)}+\ldots\B^{(2n+1)}+\ldots,
\end{eqnarray}
where superindices stand for the exponent in $\theta$ in the series
expansion. In accordance with experiments, $E$ is always a lower
order of magnitude that the magnetic field.

Similarly to a ferromagnetic media, the system might evolve to a
stable state with permanent magnetic and/or electric field because
the previous expansion not necessarily converges.

Previous mechanism, therefore, is a possible candidate for an
alternative explanation to the dynamo mechanism of the primordial
magnetic field observed in universe.

\section{Conclusions and outlook}

We have explored phenomenological consequences of Lorentz and CPT
symmetries violation through the so called non commutative field
theory.

For complex scalar field we have shown how quantum noncommutative
field theory can be constructed as a superposition of anisotropic
harmonic oscillators. Quantum fermionic fields can be constructed in
a similar way and will be reported in a forthcoming paper
\cite{lastt}.

The common feature of these approaches is their particle-antiparticle
asymmetry manifested in dispersion relations. We have explored
the possibility of using this as a possible mechanism to generate
baryon-antibaryon asymmetry as well as neutrino anti-neutrino. In the
firs case, we have shown how that can be compatible with thermal
equilibrium scenario.

For neutrinos, this asymmetry is also the mechanism that allow
flavor oscillations and then, two apparently disconnected problems
could be explained by the same mechanism. In other words, if CPT and
Lorentz are violated, would be possible, in principle, to calculate
the excess of neutrinos respect antineutrinos in the universe.

In the gauge field sector, noncommutative fields offers an
alternative process to the dynamo mechanism in order to explain
interstellar magnetic fields.

In conclusion, a tiny Lorentz symmetry violation open doors to
explain observations which can not be accommodate in the conventional
physics. Noncommutative fields, by other hand, is a description that
incorporate this properties in economical way, that is, it depends on
a small number of parameters and, just at level of free theory
exhibits features that might explain these phenomena.

\acknowledgments

This work was supported in part by US DOE Grant number
DE-FG-02-91ER40685,  FONDECYT-Chile grants 1050114, 1060079 and
2105016 and a FULBRIGHT grant (JL).


\begin{thebibliography}{99}
\bibitem{effective} For a review see {\it e.g.} J. Kogut and K. Wilson, {\it Phys. Rept.} {\bf 12}, 75 (1974).
\bibitem{yennie} See for example, C. Itzikson and J. B. Zuber, {\it Quantum Field Theory}, (1980)

\bibitem{nos2a}J.M. Carmona, J.L. Cortes, J. Gamboa
and F. Mendez; {\it JHEP} {\bf 0303}, 058 (2003).


\bibitem{nos2}J.M. Carmona, J.L. Cortes, J. Gamboa
and F. Mendez {\it Phys. Lett. } {\bf B565}, 222 (2003).

\bibitem{nos5}J. M. Carmona, J. L. Cortes, A. Das, J.
Gamboa and F. Mendez, {\it Mod. Phys. Lett. }{\bf A21}, 883 (2006).



\bibitem{gl} J. Gamboa and J.  Lopez-Sarrion, {\it Phys.Rev.}{\bf D71}, 067702 (2005).
\bibitem{jackiw} S. Carroll, G. B. Field and R. Jackiw, {\it Phys. Rev.} {\bf D41}, 1231 (1990); A. A. Andrianov, P. Giacconi and
R. Soldati, {\bf JHEP} {\bf  0202}, 030 (2002).



\bibitem{nos3} H. Falomir, J. Gamboa, J. Lopez-Sarrion, F. Mendez and A. J. da Silva, {\it Phys. Lett.} {\bf B632}, 740 (2006)
and ibid {\it Phys. Rev.} {\bf D74}, 047701 (2006).
 \bibitem{glp}J. Gamboa, J. Lopez-Sarrion and A. P.
Polychronakos {\it Phys.Lett.} {\bf B634}, 471 (2006).

 \bibitem{8}M. Gasperini, {\it Phys. Rev.} {\bf D38},
2635; S. L. Glashow, A. Halprin, P. I. Krastev, C. N. Leung and J.
 Pantaleone, {\it Phys. Rev.} {\bf D56}, 2433 (1997); R. Foot, C. N.
Leung and O. Yasuda, {\it Phys. Lett.} {\bf B443}, 185 (1998).

\bibitem{9} S. Coleman and S. L. Glashow, {\it Phys.
Lett.} {\bf B405}, 249 (1997).

\bibitem{neutrino}For a
recent review on neutrino data and experiments, see
for example A. Strumia and F. Vissani , {\it hep-ph/0606054} and
references therein.

\bibitem{more} G. L. Fogli, E. Lisi, A. Marrone and G. Scioscia, {\it
Phys. Rev.} {\bf D60}, 053006 (1999); B. Pontecorvo, {\it J. Exp.
Theor. Phys. (USSR)}, {\bf 34}, 247 (1958); L. Wolfenstein, {\it
Phys. Rev.} {\bf D17 }, 2369 (1978); S. P. Mikheyev and A. Yu.
Smirnov, {\it Sov. J. Nucl. Phys.} {\bf 42}, 913 (1985).

\bibitem{lsnd} LSND collaboration, C. Athanassopoulos et al., {\it
Phys. Rev. Lett.} {\bf 81}, 1774 (2003); LSND collaboration, A.
Aguilar et al. {\it Phys.
Rev.} { \bf D64}, 112007 (2001).

\bibitem{mb} MiniBooNE collaboration homepage:www-boone.fnal.gov

\bibitem{steril}See S. L. Glashow, {\it Phys. Lett.} {\bf B256}, 255
(1991); J. W. F. Valle
and D. Tommasini and J. T. Peltoniemi, {\it Phys. Lett.} {\bf B298},
383 (1993).

\bibitem{colgla}S. Coleman and S. L. Glashow, {\it Phys. Rev. } {\bf
  D59}, 116008 (1999).

\bibitem{kostel}V. A. Kostelecky and M. Mewes in {\it Phys. Rev.}
{\bf
  D69}, 016005 (2004), T. Katori, A. V. Kostelecky and R. Tayloe,
hep-ph/0606154.

\bibitem{last} P. Arias, Ashok Das, J. Gamboa, J. Lopez-Sarrion and
F. Mendez, {\it e-Print Archive: hep-ph/0608007 }



\bibitem{sakha}A. D. Sakharov, {\it JETP Lett} {\bf 6}, 24 (1967).


\bibitem{lastt}P. Arias, Ashok Das, J. Gamboa, J.
Lopez-Sarrion and F. Mendez, {\it in preparation}

\bibitem{sto} L. Stodolsky, {\it Phys. Rev.} {\bf D36}, 2273 (1987);
L. Stodolsky,
preprint MPI-PAE/PTh 33/88.

\bibitem{others}V. A. Kostelecky and M. Mewes in {\it Phys. Rev.}
{\bf
  D69}, 016005 (2004), T. Katori, A. V. Kostelecky and R. Tayloe,
hep-ph/0606154.




\end{thebibliography}
\end{document}